\documentclass[]{spie}  

\usepackage{amsmath,amsfonts,amssymb}
\usepackage{graphicx}
\usepackage{combelow} 
\usepackage[colorlinks=true,allcolors=blue]{hyperref}

\title{Semi-automated single-mode fiber coupling for AO-fed spectrographs}

\author[a,b]{Aaron J. Crepp}
\author[a]{Brian L. Sands}
\author[a]{Justin R. Crepp}
\affil[a]{University of Notre Dame, IN, USA}
\affil[b]{Penn High School, IN, USA}

\begin{document}
\maketitle

\begin{abstract}
   Conventional astronomical spectrographs are seeing-limited and commonly use multimode fibers to inject starlight into downstream optics. Forthcoming Doppler radial-velocity (RV) instruments are expected to make increasing use of adaptive optics and smaller single-mode fibers (SMF), enabling higher spectral resolution and improved stabilization of the spectrograph point-spread function. However, coupling starlight into SMFs with characteristic diameters of only $\sim5-10 \; \mu\mathrm{m}$ at near-infrared wavelengths is technically demanding, requiring alignment procedures that are both efficient and stable throughout the duration of an observation. In this paper, we describe a semi-automated SMF coupling control program being developed for diffraction-limited RV spectrographs. Representative simulations demonstrate that the controller reduces the star–fiber separation from approximately 3.9 mode field diameters (MFD) to 0.11 MFD while increasing a normalized coupling metric from zero to values near 0.6–0.8 after acquisition. Application to closely-separated binary stars is also studied. 
\end{abstract}

\keywords{single-mode fibers, adaptive optics, Doppler spectroscopy, radial velocity, fiber coupling, astronomical instrumentation}

\section{Introduction}\label{sec:intro}

\begin{figure}
    \centering
    \includegraphics[width=0.5\linewidth]{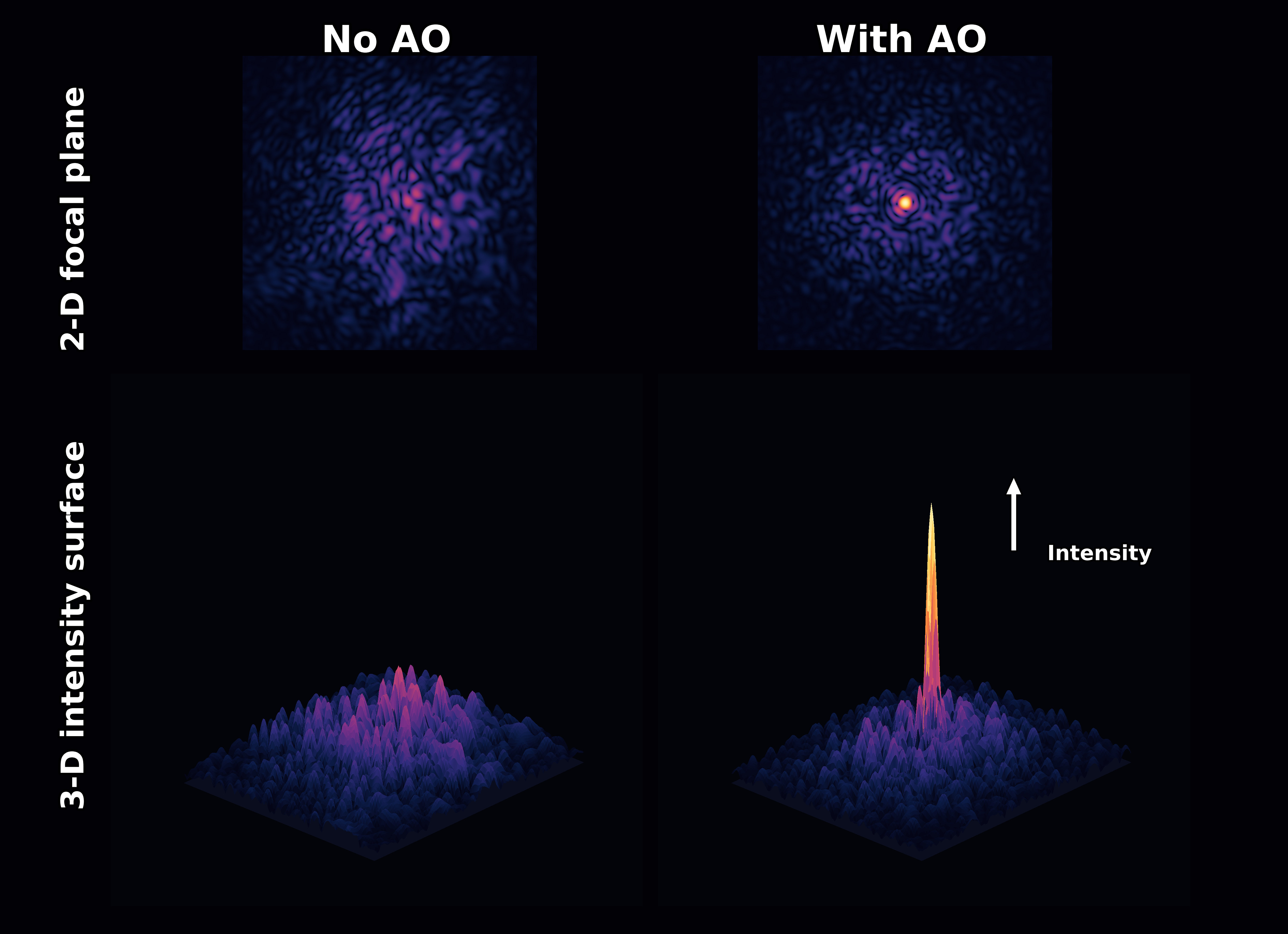}
    \caption{Example atmospheric turbulence simulation ($D/r_0=11$) showing the difference in concentrated beam power with AO off versus AO on. Uncompensated wavefront distortions distribute energy away from the Airy pattern core into the image periphery, degrading resolution and reducing peak intensity. The uncorrected image (left) reaches a normalized on-axis intensity of only 0.029, whereas the AO-corrected image (right) reaches a normalized on-axis intensity of 0.496. This dramatic ($17\times$) improvement in peak irradiance motivates the use of AO systems for Doppler spectroscopy.}
    \label{fig:AO}
\end{figure}

Modern spectrographs are now able to benefit from the use of adaptive optics (AO) (Fig.~\ref{fig:AO}) \cite{crepp_2014}. Correcting for Earth's turbulent atmosphere confers a number of benefits compared with seeing-limited spectrographs, including: higher spatial and spectral resolution; the use of single mode fibers (SMFs) to eliminate spatial modal noise; a compact opto-mechanical design leading to better stability; and lower susceptibility to OH-emission \cite{crepp_2016,bechter_2020,crass_2021}. Combined with the collecting area of a large telescope, AO-fed Doppler spectrographs show promise for stellar characterization and the detection of exoplanets using extremely precise radial-velocity (RV) measurements \cite{johnson_2026,tala_2026}.  

Although the use of AO may offer a number of advantages, in practice, time can be lost in overheads when locking control loops onto the star, finding the small ($\sim$5-10 $\mu$m diameter) SMF, and coupling starlight into the fiber. When performed manually, this process can consume valuable time at the telescope. Automating the fiber coupling sequence is, therefore, essential for maximizing scientific output over the duration of an observing program and especially over the lifetime of an instrument. 

In this paper, we describe initial efforts to develop semi-automated methods for detecting and tracking stellar targets, guiding light into a SMF, and maintaining alignment. By minimizing human-in-the-loop interactions, it is possible to recover on-sky time, better optimizing observing efficiency while minimizing downtime and mistakes. We perform numerical simulations that model atmospheric turbulence, AO correction, reference fibers, and SMF injection to demonstrate the technique and estimate coupling efficiencies. As a more challenging scenario than single stars, we also explore SMF coupling with the individual components of closely separated binaries, which represent a unique exoplanet detection science case that only AO-fed spectrographs can address. 

\section{Methods}\label{sec:methods}

\begin{figure}
    \centering
    \includegraphics[width=0.95\linewidth]{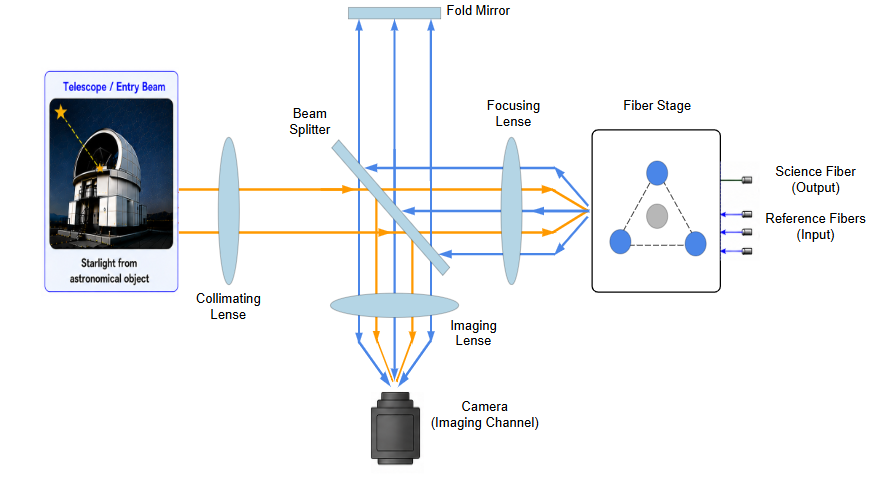}
    \caption{Conceptual layout for a SMF coupling system. Light from a star is imaged onto a camera for guiding and also passes through a beam-splitter or dichroic to be injected into the spectrograph using a SMF. Off-axis reference SMFs are back-illuminated to create an asterism (triangular pattern in this case) on the camera using the same beam-splitter and a fold mirror.}
    \label{fig:layout}
\end{figure}

We model a system similar to the optical layout shown in Figure~\ref{fig:layout}. Starlight entering from the telescope is sent to an acquisition camera where the beam is split by a beam-splitter or dichroic. Most of the photons used for science are sent to a SMF for coupling into the spectrograph, whereas some remaining light is sent to a acquisition camera that both (i) monitors image quality, and (ii) compares the position of the star with that of several reference images. In an effort to mimic existing AO-fed instruments, the reference images are produced by several off-axis SMFs that are back-illuminated by a bright source \cite{bechter_2020,crass_2021}. Light from the reference SMFs passes back through the beam-splitter to form an image on the acquisition camera. 

We generate stellar images by modeling atmospheric turbulence using Fried's parameter with a given telescope size ($D/r_0$). Rather than explicitly modeling a full AO system, for simplicity, atmospheric aberrations are partially suppressed by artificially increasing $r_0\rightarrow r_0'$. While this approach does not capture the nuance of phase correction using a deformable mirror, it provides an effective variable for dialing up/down image quality. Alongside the star, the three off-axis reference fiber images are generated using phase screens having different seed randomizations to mimic semi-static aberrations. Thus, while the stellar image varies dynamically (each frame), the reference fiber aberrations are distinct and non-evolving on the timescale of an observing sequence.

The images are collected onto the same focal plane camera array. Read-noise is injected to emulate low-level background intensity variations from the camera. Hot pixels are simulated and are identified and corrected later by the algorithm during processing. In addition to adjusting the tip/tilt of the star, it is also possible to translate the entire fiber-stage, which contains the (unseen) science fiber located near the middle of the asterism. From the perspective of a user recording data with the acquisition camera, the fiber-stage is able to move macroscopic distances across the focal plane array. In preparation for eventual on-sky testing, we assume an operational requirement that the fiber-stage must be translated, e.g. in addition to beam steering, to help achieve optimal alignment between the telescope and acquisition camera.

The aforementioned simulation code was developed in MATLAB to create realistic images as FITS files. For eventual code deployment, Python was selected to house the main logic of object detection, identification, and control input generation, as it has a wide variety of existing and helpful libraries. In addition to the main logic, the project had to also be capable of simulating the control system that would move the fiber-stage, as well as the web interface that the user would be able to monitor and interact with. For simplicity, the two supplementary programs were also developed in Python. The three independent projects run asynchronously alongside MATLAB image generation, sending data efficiently and without loss in order to make the transition to real-world applications more effective. In the interest of realism, a maximum speed was placed on the movement of the fiber-stage. 


\subsection{Detection}\label{sec:detection}

\begin{figure}
    \centering
    \includegraphics[width=0.95\linewidth]{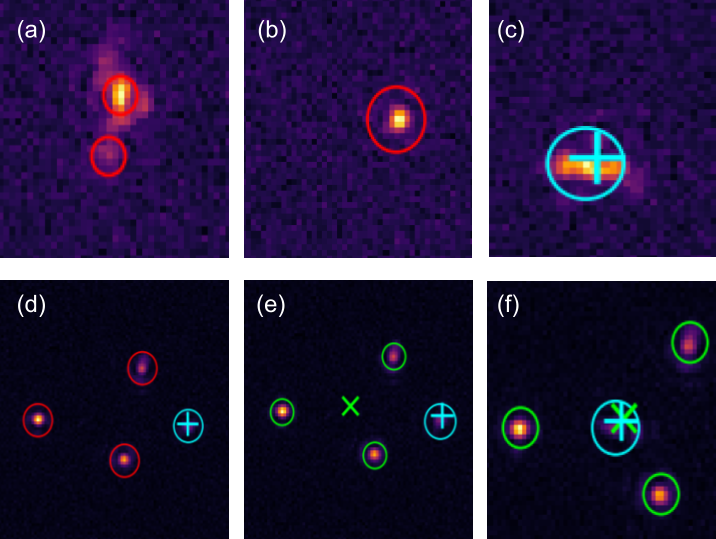}
    \caption{The six major steps for user-interaction and algorithm control. The semi-automated alignment sequence includes: (a) initial candidate detections with the reference fibers turned off; (b) detections after adjustment of the intensity and radius thresholds; (c) selected stellar target and temporally filtered centroid, marked by the cyan cross; (d) image after illumination of the three reference fibers; (e) identified reference spots and inferred science-fiber location; and (f) reference asterism after the fiber-stage has been translated to align the science-fiber location with the selected star. Each of the six images are \(256 \times 256\) pixels.}
    \label{fig:visual_walkthrough}
\end{figure}

We assume that the target star lands within the field of view of the acquisition camera system, which is on the order of several arcseconds across. Since both the star and reference fibers appear on the image as approximately Gaussian, we use a blob-finding algorithm to identify candidate science targets for the user to select. Blob-detection is implemented using the scikit-image library\footnote{\url{https://scikit-image.org/}} from Python, which employs the Laplacian of Gaussian (LoG) method. The program requires a minimum intensity threshold, as well as minimum and maximum Gaussian standard deviations ($\sigma$) of the blobs, where $\sqrt{2} \sigma$ is approximated as the blob radii in pixels. The result of blob detection is the position of each blob in pixels as well $\sigma$, which can be converted back into radius. These two measurements (position and radius) can then be used to find a brightness estimate for further characterization. The brightness estimate represents the average intensity of every pixel within the circle formed by the radius and position already found. Peak intensity is also a valid metric; however, average intensity was chosen because it offers lower fluctuations over time. These three variables (position, radius, and brightness) are consistently used throughout the program's runtime to track objects and are later used to categorize blobs as stars, reference lights, or neither. To begin the process of detecting stars, the reference fibers are turned off and the default minimum intensity and $\sigma$ parameters are used for initial blob detection. On the user-interface, any unrecognized blobs that fit the detection parameters are overlaid with a red circle and --- because at the beginning of the program nothing is recognized --- all candidate blobs are marked with red on the user-interface (Fig.~\ref{fig:visual_walkthrough}a).


\subsection{Settings and Interaction}\label{sec:settings}

The default settings are potentially imperfect, which is why some user-interaction may be necessary to assist the algorithm. The user may manually increment or decrement both minimum intensity and radius thresholds, which prevents the blob-detection algorithm from misidentifying objects amid noise (Fig.~\ref{fig:visual_walkthrough}b). Once the user is satisfied with the values they have chosen, they select a star, which in turn commences the process of finding the stellar centroid. 


Multiple images within a sequence are used to improve centroiding and brightness statistics. For each image, the stellar centroid is calculated as the intensity-weighted mean pixel position within the selected blob. The resulting centroid estimate is appended to a running history, from which a more stable position is determined using an exponentially weighted average that assigns greater weight to more recent measurements to accommodate systematic image drifts. This temporally filtered centroid is displayed as a cyan ``+'' (Fig.~\ref{fig:visual_walkthrough}c). Once the desired star has been reliably identified and centered, the user may finalize the minimum radius and intensity thresholds, ending the interactive portion of the procedure.

\subsection{Reference-spot Identification}\label{sec:identify}

Once the user is finished with selecting a star, the three reference fibers are turned on and identified within the image (Fig.~\ref{fig:visual_walkthrough}d). To help distinguish reference SMFs from science targets, it is useful to take an image before turning on the reference fibers and subtract that image from one where the fibers are illuminated. To reduce residuals caused by temporal variability in the stellar images, several frames are averaged before and after illuminating the reference fibers, and the two averaged images are then subtracted.


Once the differenced image is created, the blob-detection algorithm should only detect the three reference lights and their metrics of intensity, position, and radius can be recorded and tracked. In the case where less than three blobs are detected, the star-subtraction process can be rerun, a separate identification algorithm can potentially be employed, or the program can end in failure as the preconditions of all three reference lights being on the image may not be met, as decided before the program is run. In the case where more than three blobs are detected, a cost function is employed to identify which blobs correspond to the reference fibers. The algorithm searches every unique combination of three blobs, punishing each based on uniformity of distance from each other (as the reference lights form an equilateral triangle), radius, and intensity. When the reference fibers are found, they are identified with lime circles on the web interface using a lime ``X" at the center of the three (Fig.~\ref{fig:visual_walkthrough}e).

\subsection{Fiber-stage Control and Tracking}\label{sec:control}
With estimates of radius and intensity as well as initial positions of each star and reference fiber, it is then possible to begin translating the reference fibers to the time-averaged centroid of the chosen star (Fig.~\ref{fig:visual_walkthrough}f). This movement is achieved through a proportional-integral (PI) controller. The proportional term represents a measurable gain value with appropriate units that sets how aggressively the fiber-stage movement algorithm closes in on the star. A small integral term may be useful in the presence of driver-motor friction in practice. 



The reference fibers must be tracked during the movement process to provide feedback for the PI loop. Because the reference fibers have relatively stable aberrations, the radius and intensity measurements are assumed to be constant throughout the runtime of the program. Tracking the fibers is done through a predictor-corrector scheme. The fiber position prediction uses the explicit Euler method with discrete time steps using the output of the PI controller as the velocity estimate (which can be tuned based on the physical speed/acceleration limits of the fiber-stage). This prediction is used as the position that is fed into the PI controller. In parallel with the prediction, a secondary position estimator --- that does not take into account the predicted positions --- instead finds the objects that most resemble the stars recorded, adding to a total cost metric if the stars have a difference in intensity, radius, or position from their originally recorded values (which is helpful when the star and reference fiber overlap). If there are three objects remaining, it finds a cost value in a similar way to the process described in the explanation of Fig.~\ref{fig:visual_walkthrough}e of the visual walkthrough and adds it to the cost metric. This cost must be less than a predetermined value for the original prediction to be overruled. The purpose of this multistage tracking system is to help protect against common edge-cases that are present in the problem. Because the predictions are based on the output of the PI controller and this output is based on the position prediction for the reference lights, the system needs real feedback from the images. If a star and reference fiber overlap, the second algorithm can fail, which demonstrates the purpose of the prediction as a metric available every frame.

\subsection{Coupling Calculation}\label{sec:coupling}
\begin{figure}
    \centering
    \includegraphics[width=0.95\linewidth]{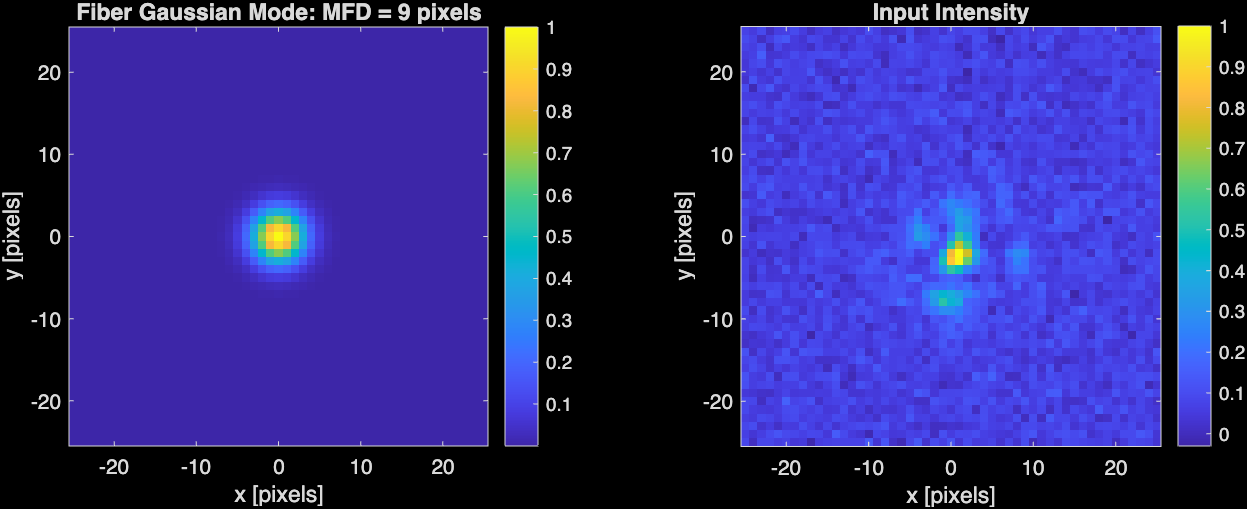}
    \caption{SMF coupling simulations for a single star. (Left) SMF mode field and (Right) stellar image shown on a normalized intensity scale. Coupling efficiency is estimated using an overlap integral between the fiber mode field and image.}
    \label{fig:mfd}
\end{figure}

To estimate fiber coupling efficiency, we calculate the overlap integral of the stellar complex field with the Gaussian mode field of the SMF,
\begin{equation}
    \eta = \frac{|\int \int E_{\rm star}(x,y) E^*_{\rm SMF}(x,y) dx dy|^2}{(\int \int |E_{\rm star}|^2 dx dy)(\int \int |E_{\rm SMF}|^2 dx dy)},
\end{equation}
where the integrals are evaluated over a sufficiently large range (numerous mode field diameters (MFDs)) to enable convergence. As the star experiences different turbulence realizations from the atmosphere and AO system correction, we resolve the time-series using data from the simulated acquisition camera. It is assumed that the SMF produces an effective mode field diameter of 9 pixels (arbitrarily chosen) as projected onto the focal plane array after magnification. Figure~\ref{fig:mfd} shows an image of the SMF Gaussian mode field alongside a representative AO image. The detection, steering, and pointing maintenance algorithm is used to optimally align target stars to the spectrograph SMF using the reference images in an effort to optimize the overlap integral. 

\section{Results}\label{sec:results}

We characterize SMF coupling based on the semi-automated algorithms presented in $\S$\ref{sec:methods}. Star-to-fiber distance and injection efficiency are quantified for AO data sets involving both single stars ($\S$\ref{sec:single}) and binary stars ($\S$\ref{sec:binary}). In the latter case, we also quantify unwanted contamination from a closely-separated companion star. 

\subsection{SMF Coupling with Single Stars}\label{sec:single}

\begin{figure}
    \centering
    \includegraphics[width=0.95\linewidth]{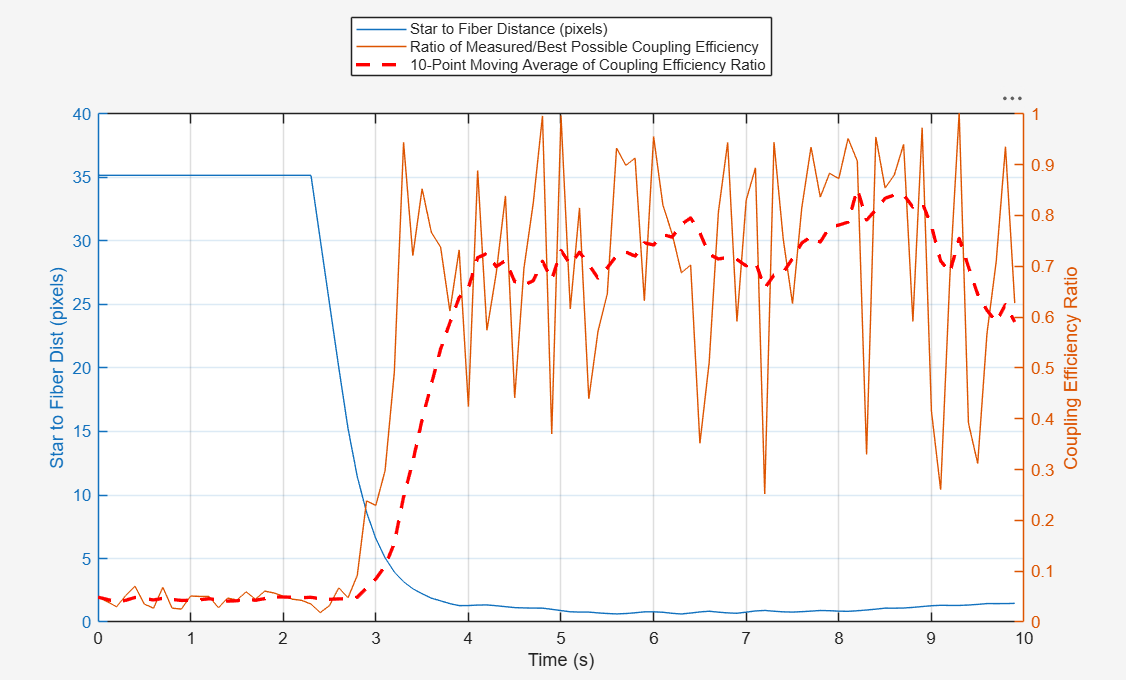}
    \caption{Closed-loop single-star acquisition in a representative simulation. The blue curve shows the separation between the selected stellar centroid and the inferred science-fiber position. The orange curve shows the instantaneous normalized coupling metric, $R(t)=\eta(t)/\hat{\eta}_{\max}(t)$, where $\eta(t)$ is the overlap at the commanded fiber position and $\hat{\eta}_{\max}(t)$ is the largest overlap found by scanning trial fiber positions around the instantaneous stellar PSF. The red dashed curve shows a 10-sample moving average of $R(t)$. After target and reference-fiber identification is completed at $t\approx2.4$~s, the PI controller reduces the star--fiber separation to approximately one pixel while the smoothed normalized coupling metric rises toward $\sim$0.6--0.8. The time axis is only illustrative because the simulated stage dynamics and controller gains were not calibrated to a specific instrument.}
    \label{fig:coupling}
\end{figure}

Figure~\ref{fig:coupling} shows a representative closed-loop acquisition sequence for a single star. Because residual atmospheric aberrations change the instantaneous stellar point-spread function (PSF), the maximum achievable coupling also varies from frame to frame. We therefore normalize the coupling estimate at the commanded fiber position, \(\eta(t)\), by an estimated instantaneous maximum,
$$ R(t)=\frac{\eta(t)}{\hat{\eta}_{\max}(t)}. $$
The quantity \(\hat{\eta}_{\max}(t)\) is obtained by evaluating the same overlap metric on a \(25\times25\)-pixel grid of trial fiber positions around the star and retaining the largest value. Thus, \(R=1\) indicates that the commanded position attains the best coupling found for that instantaneous PSF; it does not indicate 100\% absolute coupling.

Prior to \(t\approx2.4\) s, the program is identifying the selected star and the three reference-fiber images, while the simulated fiber-stage remains stationary. Once identification is complete, the PI controller is enabled and the star--fiber separation decreases rapidly from approximately 35 pixels ($3.9$ MFD) to about one pixel ($0.1$ MFD). The instantaneous value of \(R(t)\) continues to fluctuate as the stellar PSF varies between frames. Nevertheless, its 10-sample moving average rises from its initial value and settles near $\sim$\(0.6\)--\(0.8\), which demonstrates the basic capability of the fiber-coupling algorithm for acquiring a simulated target star. The normalized metric remains below unity even after the centroid error becomes small. One possible explanation is that the intensity centroid of an aberrated PSF does not necessarily coincide with the position that maximizes coupling into the SMF. Controller latency, detector noise, stage-position sampling, and the finite grid used to estimate \(\hat{\eta}_{\max}\) also contribute. The simulation demonstrates convergence of the alignment loop, but the horizontal time scale should not be interpreted as an acquisition-time (wall-clock) prediction for a specific instrument. The stage-speed limit, update rate, and controller gains were selected for the simulation rather than derived from measured hardware. Laboratory testing will be required to determine absolute acquisition time, steady-state alignment error, repeatability, and end-to-end coupling efficiency.

\subsection{Application to Binary Stars}\label{sec:binary}

AO-fed spectrographs are uniquely capable of studying planets located in binary star systems (s-type orbits). By isolating the light from individual component stars, it becomes possible to study planet statistics as a function of binary separation and mass ratio. The resulting occurrence rates, orbital eccentricity and orientation, as well as other parameters, can help to inform planet formation and evolutionary theories, providing complementary data to studies of single stars alone \cite{gardner_2022}. At the same time, AO imaging can resolve stars previously thought to be single, or systems already known to host planets from other methods such as transits \cite{crepp_2025,collier_2026}. 

\begin{figure}[t]
    \centering
    \includegraphics[width=0.60\linewidth]{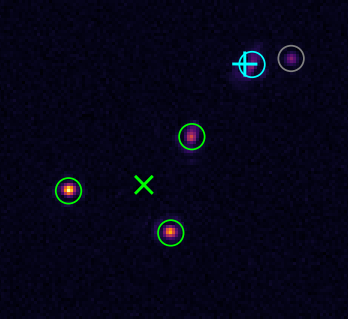}
    \caption{Example binary star simulation showing the reference fibers before alignment. The science fiber is indicated by an `X' and the primary star indicated by a `+' sign.}
    \label{fig:binary}
\end{figure}

\begin{figure}[h]
    \centering
    \includegraphics[width=0.71\linewidth]{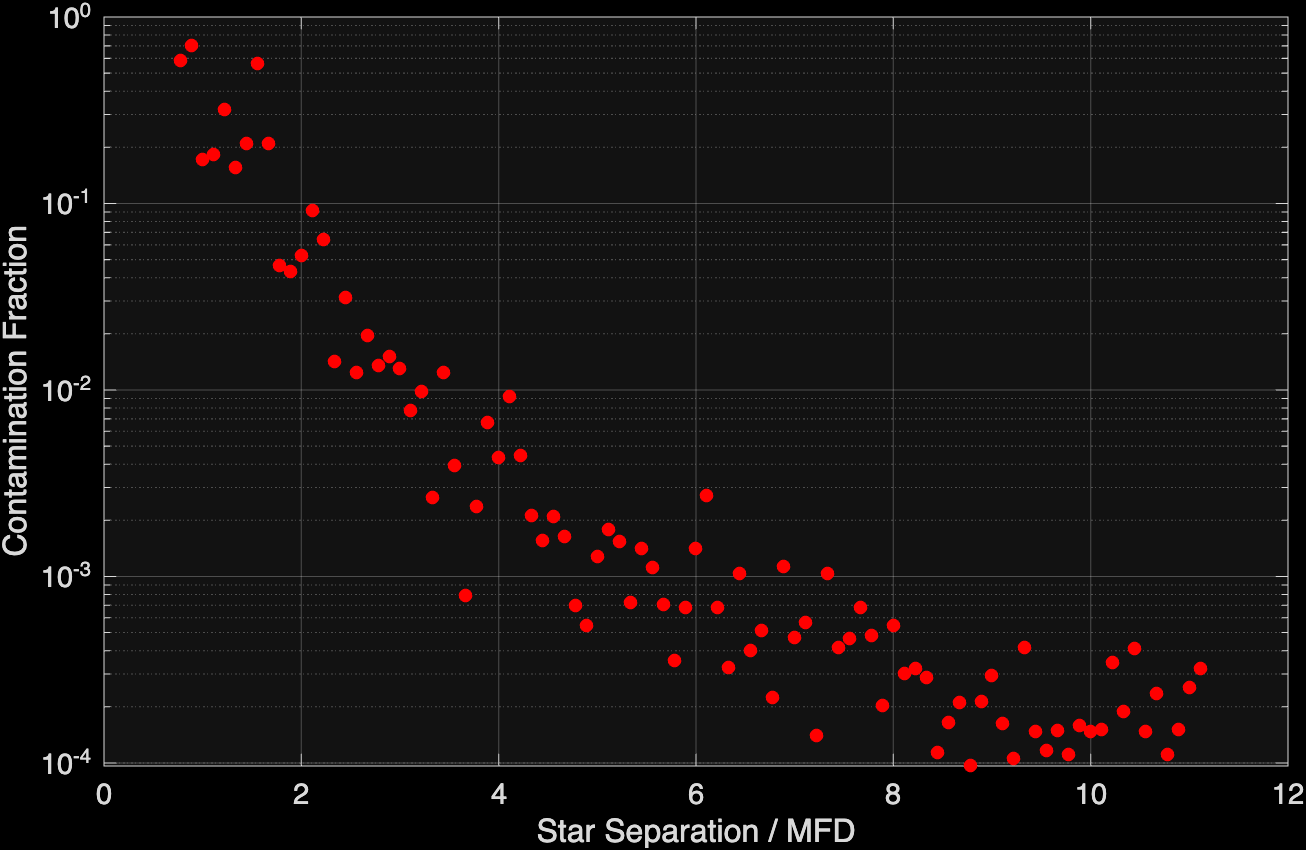}
    \caption{Fractional SMF contamination from a nearby binary star of equal brightness. Each point represents an average of 15 atmospheric realizations. The angular separation is normalized by the MFD.}
    \label{fig:contamination}
\end{figure}

We estimate the time-averaged cross-contamination, $<(F_{\rm comp}\eta_{\rm comp})/(F_{\rm target}\eta_{\rm target})>$, caused by a nearby star when targeting the individual component of a binary (Fig.~\ref{fig:binary}). It is assumed that the binary stars have equal brightness, $F_{\rm comp}/F_{\rm target} = 1$, since scaling the relative intensities can estimate contamination with different brightness ratios. Parallactic rotation is neglected over the short simulated acquisition sequence; its effect during longer observations will be incorporated in future pointing-maintenance simulations. Although the algorithm can be run throughout a longer observation, pointing maintenance and further optimization is assumed to be handled using feedback from a receiver that monitors SMF injection in real-time, which can only be run once a star is coupled with the fiber. 

Figure~\ref{fig:contamination} shows results for fractional contamination as a function of binary star separation. The simulations averaged the results of 15 different atmospheric realizations for each separation. We normalize the separation based on the SMF MFD as projected onto the acquisition camera. The contamination fraction is found to fall off rapidly with distance due to the high image quality provided by the simulated AO system, with Strehl ratios routinely exceeding 50\% in the near-infrared. We note that the fractional contamination is slightly biased at wider separations due to rounding errors (near-zero coupling) and finite field of view over which the overlap integral was evaluated. Another limitation is that these simulations did not explicitly model the AO wavefront sensor; certain types of sensors like the Shack-Hartmann can get confused with resolved binaries. However, pyramid sensors, such as those deployed at the LBT, offer better performance with closely-spaced guide stars.  

To first order, the Doppler uncertainty induced by a nearby star is proportional to the contamination fraction (though depends on spectral type, relative RVs, rotational broadening, data reduction method, and other variables in practice). For many RV spectrographs, a measurement uncertainty of $10^{-3}$ pixels corresponds to roughly 1 m/s precision (to within an order of magnitude) in the absence of other systematics. Equating these levels to the SMF contamination results suggests that a diffraction-limited spectrograph could achieve $\sim$1 m/s precision as close as $\approx 5$ MFDs away from the primary star in the case of an equal-brightness binary. This level of performance however needs to be confirmed with on-sky measurements. 


\section{Summary and Conclusions}\label{sec:summary}

We have developed a semi-automated program for acquiring, detecting, and guiding starlight into SMFs using image data from an acquisition camera combined with back-illuminated off-axis reference fibers. The reference fibers form an asterism that may be turned on/off and is steerable. With minimal user-interaction required, e.g. only the confirmation of which blob to target, we find that the control-loop program converges to approximately one-pixel alignment ($0.11$ MFD) and substantially improves the normalized coupling metric. Laboratory validation will be required to quantify absolute coupling efficiency, acquisition time, robustness, and repeatability.

We have further explored the capability of targeting the individual components in closely-spaced binary star systems. We find that the detection and guiding algorithm is able to latch onto sources selected by the user, and that contamination from equal-brightness stars is minimal beyond $\approx5$ MFDs depending on seeing-conditions. Next steps include taking into account changes in parallactic angle throughout the course of an observation and measuring wavelength-dependent cross-contamination between sources as a function of spectral-type and other variables for binary stars. 

These results are consistent with the findings of other research groups and instrument concepts, which conclude that the performance of a well-designed and well-built SMF injection system is governed primarily by AO performance \cite{schwab_2016,gibson_2023,jovanovic_2024}. Provided that systematic errors from vibrations and other effects are minimized (e.g. imperfect atmospheric dispersion correction), automated SMF coupling could become a routine and reliable part of operations for a diffraction-limited spectrograph. The software that oversees, controls, and maintains fiber coupling can be automated to the point that few interactions with the user (though perhaps not zero) are necessary to optimize and sustain consistent on-sky performance. While the methods described are currently general in nature, it is envisioned that these techniques could be deployed for eventual operation with SMF spectrographs on telescopes that use AO \cite{crepp_2016,mawet_2016,feger_2018,kotani_2020,crass_2022,vigan_2024}. Further research will be needed however to maintain optimal SMF coupling efficiency using ancillary hardware beyond just feedback from an imaging camera, such as the use of sensitive receivers that monitor injected stellar flux in real-time. 


\acknowledgments
This research was supported in part through a service agreement with Notre Dame's Engineering and Design Core Facility (EDCF).

\bibliographystyle{spiebib}
\bibliography{bibliography}

\end{document}